\documentclass[11pt,a4paper,oneside]{article}

\usepackage{graphicx}
\usepackage{topcaption}

\begin{document}

\title{Trend arbitrage, bid-ask spread and \\market dynamics}
\author{Nikolai Zaitsev\footnote{Fortis Bank, Risk Management Merchant Bank, 
Correspondent e-mail: nikzaitsev@yahoo.co.uk}
\\(submitted to "Quantitative finance" journal)}
\date{May-June 2006}

\maketitle

\begin{abstract}
Microstructure of market dynamics is studied through analysis of tick price data. 
Linear trend is introduced as a tool for such analysis. Trend arbitrage inequality is 
developed and tested. The inequality sets limiting relationship between trend, 
bid-ask spread, market reaction and average update frequency of price information. 
Average time of market reaction is measured from market data. This parameter is interpreted as 
a constant value of the stock exchange and is attributed to the latency of exchange reaction to 
actions of traders. This latency and cost of trade are shown to be the main limit of bid-ask spread. 
Data analysis also suggests some relationships between trend, 
bid-ask spread and average frequency of price update process. 
\end{abstract}

\section{Introduction}

Question about market microdynamics is of fundamental importance. It influences wide range of items: 
bid-ask spread dynamics, interaction of market agents setting market and limit orders,
patterns of price formation and provision of liquidity. 
These subjects are investigated using wide range of analytic tools, trying 
to capture cross-correlation structure of price evolution and volatilities through 
time series analysis~\cite{Roll} or studying dynamics of price formation through 
simulation of detailed market mechanics~\cite{Farmer} \footnote{Such simulations are part of 
so-called 'Experimental Finance'.}. 
These methods concentrate on search of the source of random forces acting over 
continuous or discrete time sets. These sets are considered homogeneous when long-term returns are considered.
On the shorter scale (in tick data), it is noticed that the trading frequency scales the 
returns making time sets non-homogeneous. Theoretical construction is 
built in~\cite{EDerman} to account for such effect.

In contrast to statistical methods, 
the behavioral finance searches patterns of dynamics of collective actions of various market agents  
making assumptions about their perception and searching for equilibrium 
and/or limiting solution to price/spread prediction problem~\cite{AShleifer, Bondarenko}. 

All above mentioned methods analyze price returns. In this article we introduce new object for such analysis, 
a trend~\footnote{In fact, trend was studied many times before, but in context of technical analysis and 
it was always interpreted as an evidence to some non-random (deterministic) price behavior.}. 
Common definition of trend is \textit{"the general direction in which something tends to move"}. 
Visually trend can be identified as a line with little random deviations.  Of course, trends were studied 
many times before. However, they were studied often in relation to question of deterministic behavior of price.

We think, that the existence of trend is not $necessarily$ deterministic and in general does  
not disapprove statistical nature of price behavior. Indeed, there is a non-zero 
probability to find straight line in price path generated using, for example, Black-Scholes model~\cite{RandomWalk_Trend}. 
Despite this fact, crowd of market participants still believes that such trends are due to 
some hidden bits of information and therefore contribute to trend persistence or otherwise. 

In this paper we would like to merge these two ideas of considering price evolution through scope of 
trend and through random walk. Indeed, 
the presentation of price path in terms of price returns (random walk) is suitable in connection to CAPM  
and alike theories, where profit and risks are related in search of optimal investment portfolio. 
On the other side, presentation of price evolution in terms of short-term price trends better conforms description 
of market microdynamics. 

The trend is an object worth to study because as we show it gives insight into microstructure of market dynamics. 
It allows direct observation of spread dynamics, price information update, costs, trading activity and response time 
of the exchange. This information is diluted when correlation analysis of price return series is used.
In this paper trends are studied without consideration of news impact to the price move but only 
in relation to dynamics of agents interaction via exchange. 

Exchange is a dynamic system. It allows traders to interact with each other. 
It consists of many components. These parts are specialists on the floor, 
electronic platform which helps to account and match trades through software and 
a wire connection to trader computers. Traders use electronic platforms to trade. 
Both systems, of traders and of exchange, have their respective non-zero latencies and therefore the speed of information
diffusion is non-zero~\footnote{At the end, there is a hard limit of speed of light in the transmission cables.}.
The latencies influence price dynamics and respective trading algorithms.

This study draws attention to the influence of dynamics of systems used for trading on the dynamics of price. 
We extract information content of such dynamics from trends and spreads data.

The article is organized as follows. First, we discuss trend arbitrage. 
The respective relationship has been derived and is taken as starting point to further 
analysis. Second, one applies an algorithm searching for linear trend in time series of 
bids and asks observed on stock markets. Third, several distributions are obtained to 
support the relationship developed from trend arbitrage requirement. Several outcomes 
have been discussed as well.

\section{Trend Arbitrage}
\label{sec:trend_arb}

In the following, one analyze intra-day trend seen in prices of an asset which is 
traded in double auction.

For brevity of explanation we assume two kind of market agents: 
Specialists and Investors. Specialists maintain sell- and buy-prices (asks and bids, 
respectively) in continuous manner. Bids and asks generally move together. Spread between them 
(spread=ask-bid) tends to some minimum level due to competition between Specialists. 
This level depends on risk perceived by Specialists to take their market share. 
In stress situations, especially when price directionally moves to the new level, 
spread increases. The question now is what this dependency is.

\begin{figure}[!htbp]
\topcaption{Trend arbitrage relationship.\label{fig:trend_arb}}
\includegraphics[width=0.9\textwidth]{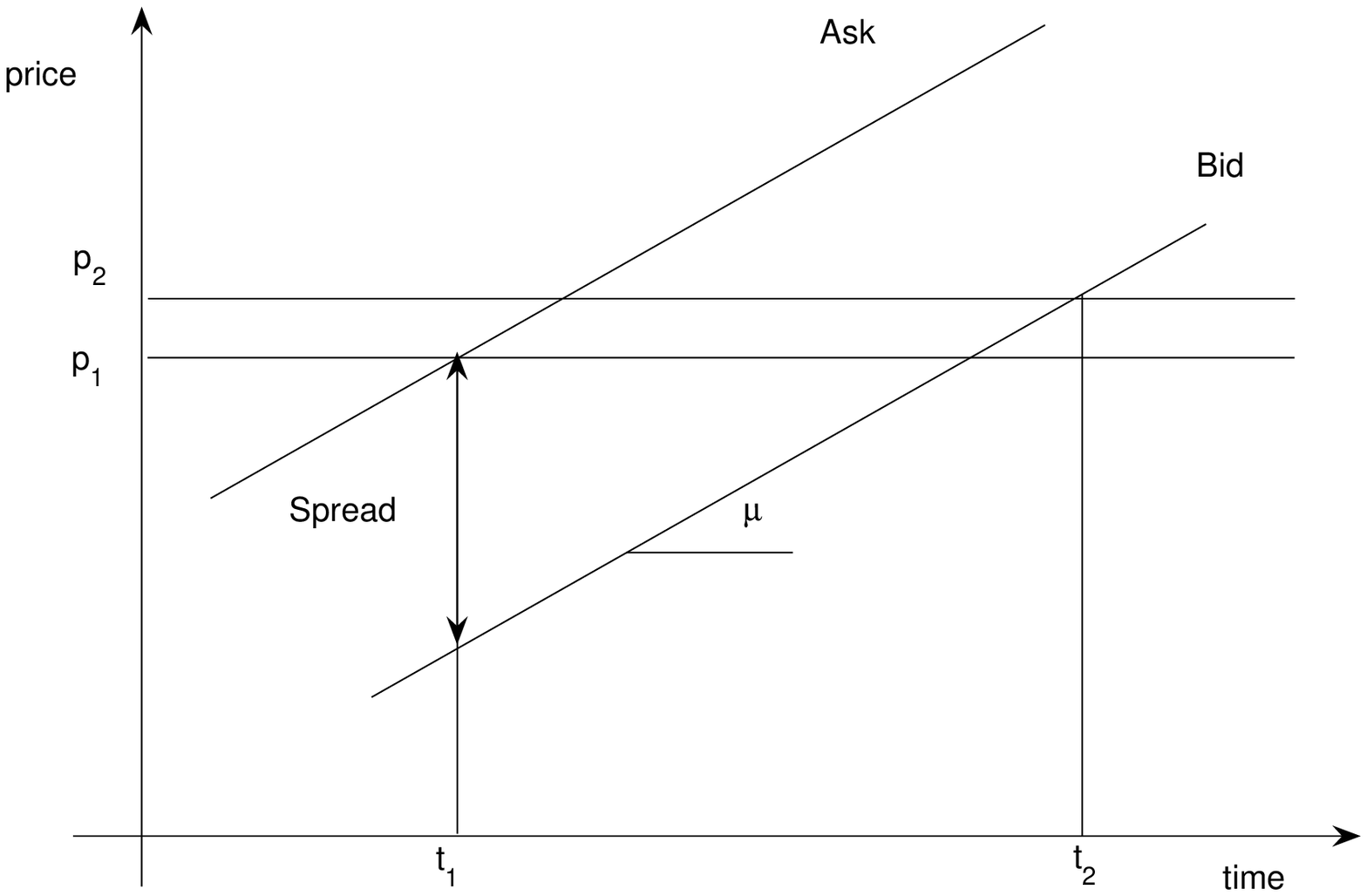}
\end{figure}

A simple example helps to find an answer. Specialist provides bid and ask prices earning spread. 
Investor profits by taking directional bets. Suppose now that 
the price moves within up-trend, $\mu$, with spread, $Spr$, see Figure~\ref{fig:trend_arb}.
\\Investor buys asset at ask-price, $p_1$, at $t_1$ moment. After a while he sells it. 
If the trend did not change its direction and Investor waits long enough, till time $t_2$, he can 
sell asset at bid-price, $p_2$. Investor will make \textit{no profit} if:

\begin{equation} \label{eq:arb1}
0 \geq p_2 - p_1 - Spr + \mu \cdot (t_2 - t_1)
\end{equation}

Investor must meet three conditions to profit for sure:
\begin{itemize}
\label{cond:trend_arb}
\item $|\mu|$ is non decreasing during the period between $t_1$ and $t_2$;
\item trading system of Investor is faster than that of Specialist and trading system of Specialist is 
slower than some market reaction value, $\tau$;
\item spread is small enough, satisfying~(\ref{eq:arb1}).
\end{itemize}

Knowing these conditions (consciously or not), Specialist sets spread such that he makes no losses, 
i.e. $p_2$ is kept equal to 
$p_1$ or less. 

Same arguments are applied to down-trend, which changes $\mu$ to $|\mu|$. 
Therefore, (\ref{eq:arb1}) is modified and limit on spread can be set:
\begin{equation} \label{eq:arb2}
Spr \geq |\mu| \cdot \tau
\end{equation}
\\where $\tau=t_2-t_1$. More general expression must also include additional costs to be paid by Investor 
related to his trading activity and some structure of characteristic time, $\tau$:

\begin{equation} \label{eq:arb3}
Spr \geq |\mu| \cdot (\tau_M + \tau_S) + fee
\end{equation}
\\where $fee$ is a broker fee and similar additive costs, $\tau_M$ is 
exchange-wide latency and $\tau_S$ is stock specific latency. $\tau_S$ can be attributed to aggregated 
latencies of trading platforms used by Investors to trade the same stock.

These inequalities suggest that the non-zero spread is due to both, transaction costs 
and latencies of trading and exchange systems.

Now, let us look at the relationships through data analysis perspective.

\section{Data analysis}

\subsection{Data}

Tick price/volume data of 33 stocks traded at Amsterdam (NLD), Frankfurt (GER), Madrid (SPA), Paris (FRA) and 
Milan (ITL) were collected over the second half of the year 2003. This data includes bid,
ask and trade prices and respective volumes attached to time stamps. Time period 
considered is between 15:30 and 17:30 of European time 
\footnote{this specific choice of time was motivated by wish to study cross-dynamics of 
European and American markets.}. 
Dividend information was not used because it does not affect our intraday analysis.

Data were filtered. Only those bits of information were used in analysis where 
price of either bid or ask has been updated. Any such change is treated as "information update".

\subsection{Trend identification}
\label{sec:trend_id}

\begin{figure}[!hbp]
\centering
\topcaption{Trend fit to bid-ask series of ABN-AMRO. 
Star-points connected with lines are bid and ask prices (in blue and red, respectively), 
piece-wise solid lines in between are fitted trends.
\label{fig:trend_fit}}
\includegraphics[width=0.75\textwidth]{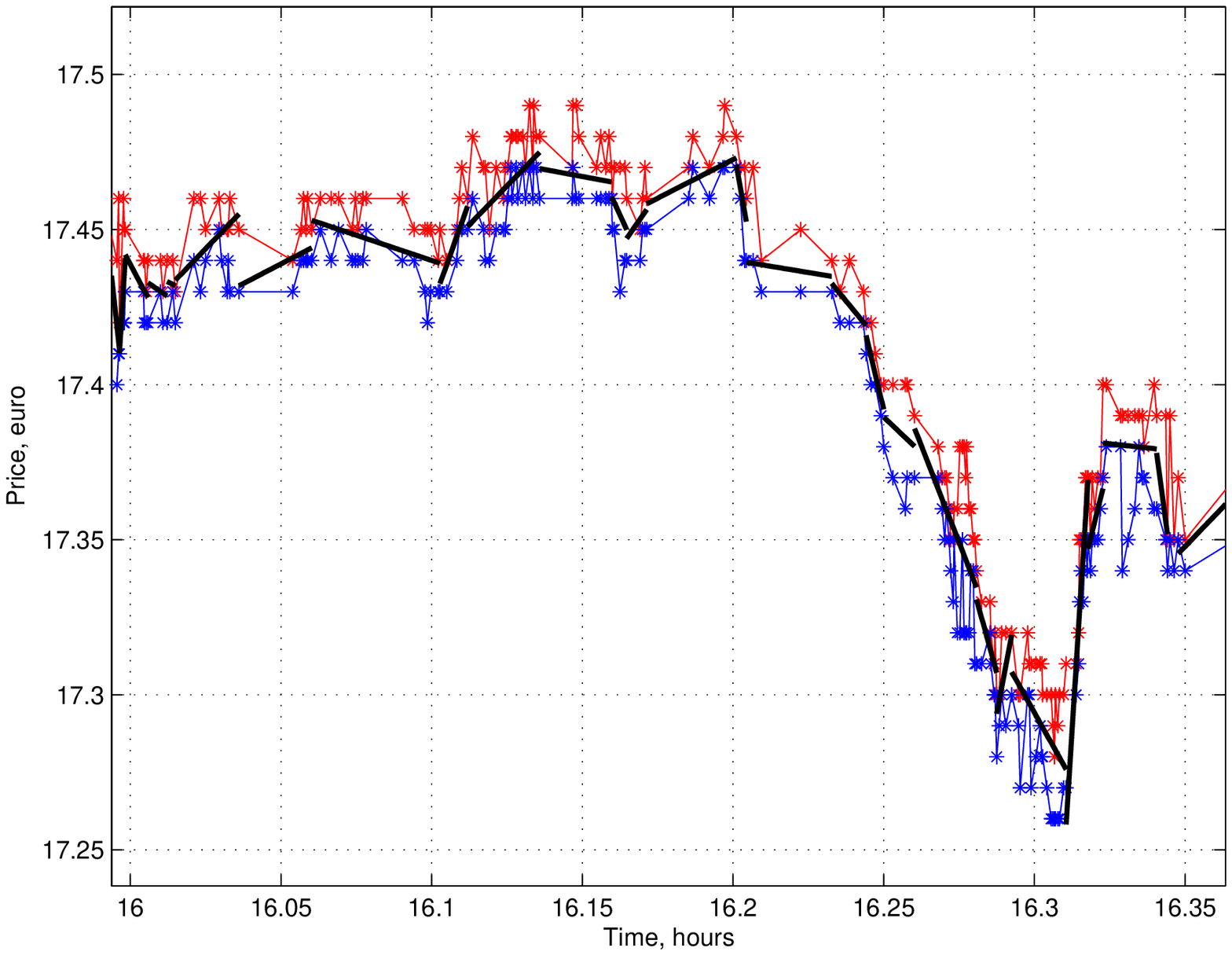}
\end{figure}

Linear fit (details can be found elsewhere \cite{Handbook}) 
was applied to series of one day mid-prices, $(ask(t_i)+bid(t_i))/2$. 
Bid-ask spread, $\epsilon(t_i)=(ask(t_i)-bid(t_i))$, is understood as an uncertainty of observed 
price information (this item to be discussed further). 
Data were fit continuously. This way real-time feed is simulated, where only past information 
is available. An output of fit is series of current trend parameters such as series 
of impact, $p_0(t_i)$, series of slope (trend), $\mu(t_i)$, with respective errors, $\epsilon_p(ti)$, 
$\epsilon_{\mu}(ti)$. 
These parameters allow prediction of forward price, $\tilde{p}$, and price error, $\epsilon$: 
\begin{equation} \label{eq:fwd_price}
\tilde{p}(t_i) = p(p_0(t_{i-1}),\mu(t_{i-1}),t_i) \textrm{, and}
\end{equation}
\begin{equation} \label{eq:e_fwd_price}
\tilde{\epsilon}(t_i)= \epsilon(p_0(t_{i-1}),\mu(t_{i-1}),\epsilon_p(t_{i-1}),\epsilon_\mu(t_{i-1}),t_i). 
\end{equation}
\\Condition to trigger end of current trend:

\begin{equation} \label{eq:trigger}
	\frac{|\tilde{p}(t_i) -p(t_i)|}{\sqrt{\tilde{\epsilon(t_i)}^2+\epsilon(t_i)^2}} > n
\end{equation}
\\where $n$ is cut value, which is chosen to be equal to 2.3. 
This particular choice of cut-off value was determined visually from trends 
superimposed on the price plot (see Figure (\ref{fig:trend_fit}). 
There is an intuitive feeling that the less cut-off value generates too many spurious trends and 
dilutes effects we are looking for. Use of price returns is a limiting case for trends 
with just two points. On the other side, the larger cut-off value will lead to reduced 
sensitivity of spread information to trends. 

\begin{figure}[!hbp] 
\centering 
\topcaption{$\chi^2$ probability for all found ABN-AMRO trends.\label{fig:prob_chi2}} 
\includegraphics[width=0.75\textwidth]{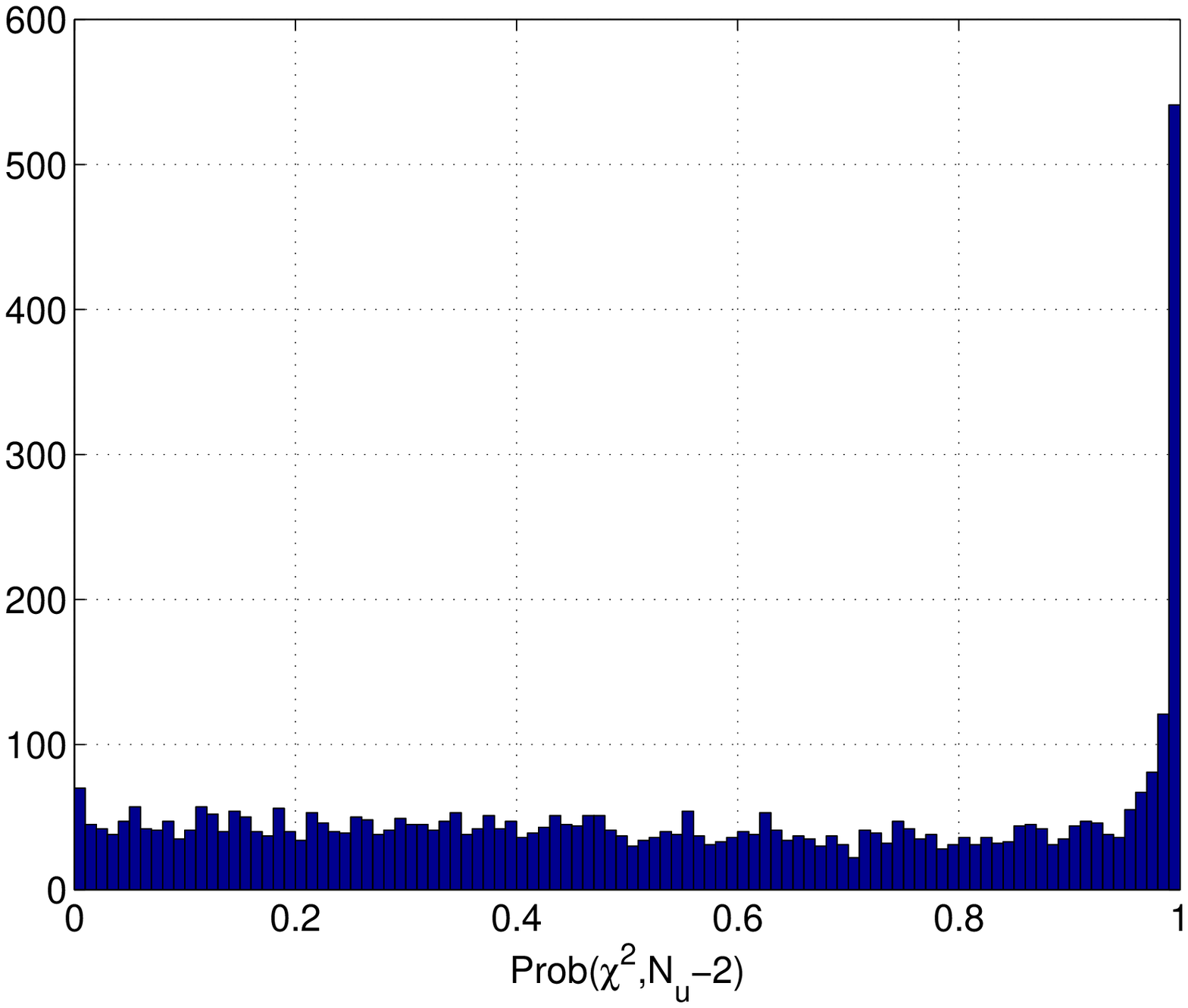} 
\end{figure}

Trend search starts at the beginning of each day series with a seed of three points. Algorithm continues 
to fit line till the trigger (\ref{eq:trigger}) is hit. It then assumes that within the search time window
(between current point and the beginning of current trend)
there are two trends. Algorithm starts backward search in order to find an optimal breaking time point 
between current and forward trends. 
The point is found at the minimum of $\chi^2_t=(\chi^2_t)_{cur} + (\chi^2_t)_{fwd}$, where $(\chi^2_t)_i$ are of current 
($i=cur$) and of forward ($i=fwd$) trends. This point becomes the end of current trend and the start of forward one. 
The forward trend becomes current. And so on till the end of day. Each trend is then accompanied with respective 
information:
\begin{itemize}
\item start and end point within time series, $t_0$ and $t_e$;
\item duration of trend in seconds, $\Delta T=t_e-t_0$;
\item slope, $\mu$, impact, $p_0$, and respective errors, $\epsilon_s$, $\epsilon_p$ 
found at end-point, $t_e$;
\item quadratic average spread, $\tilde{Spr} = \sqrt{\frac{\Sigma \epsilon^2}{n}}$;
\item number of data points in the fit, $N_u$. $\tau_u=\Delta T/N_u$ average time 
between updates of price (bid or ask) information. $\tau_u$ is also called average update time;
\item $\chi^2_t$ of the fit, where 
\begin{displaymath}
\chi^2_t = \sum_i \frac{(\tilde{p}(t_i) -p(t_i))^2}{(\epsilon_p)_i^2}
\end{displaymath}
\\Trends and associated data found from intraday data were merged into one single 
dataset day-by-day. For further analysis a selection cut was applied to filter badly fit trends:
\begin{itemize}
\item $\chi^2_t$-probability, $Prob(\chi^2_t,n.d.f.)<0.99$, where $n.d.f = N_u-2$;
\item number of points in the fit $>4$;
\end{itemize}

\subsection{Price uncertainty}

Trend measurement rises important question: what is the measurement error of observed price? 
One possibility is to interpret mid-prices as observed exactly with equal weights. 
The errors of fitted line parameters can be estimated from the sample in a standard way.
Another way is to assume that the price is governed by some diffusion process and therefore to interpret 
estimated volatility as current price errors. Both methods seem not acceptable since they do not bear instant information 
about price uncertainty.

We believe that the bid-ask spread is a good candidate to estimate instant price uncertainty. In principle, 
the price error (and spread) must be dependent on the volume to be traded by some arbitrage algorithm, which uses 
the trend fit. In this case all analysis will change.
In this research we concentrate on analysis of the best bids-asks dynamics alone. 

Example of distribution of $\chi^2_t$-probability (see Figure~\ref{fig:prob_chi2}) demonstrates that our 
choice is overall correct because the probability is mostly flattened in range [0,1]~\footnote{flat distribution of 
$\chi^2$-probability generally indicates that model errors correctly describe the measurement process.} peaking at 1. 
\end{itemize}
The rejected trends with $\chi^2_t>0.99$ tend to have smaller $|\mu|$. In case of less expressed trends the market 
enters intermediate regime known from technical analysis as "side trend".

\section{Results} 

\subsection{Test of trend arbitrage}

\subsubsection{Market reaction time, $\tau_M$} 
\label{sec:market_reaction}

\begin{figure}[!hbp] 
\centering 
\topcaption{Distribution of $\tau_{resp}$ for ABN-AMRO stock. Red line indicates log-normal fit.
\label{fig:market_response}}
\includegraphics[width=0.75\textwidth]{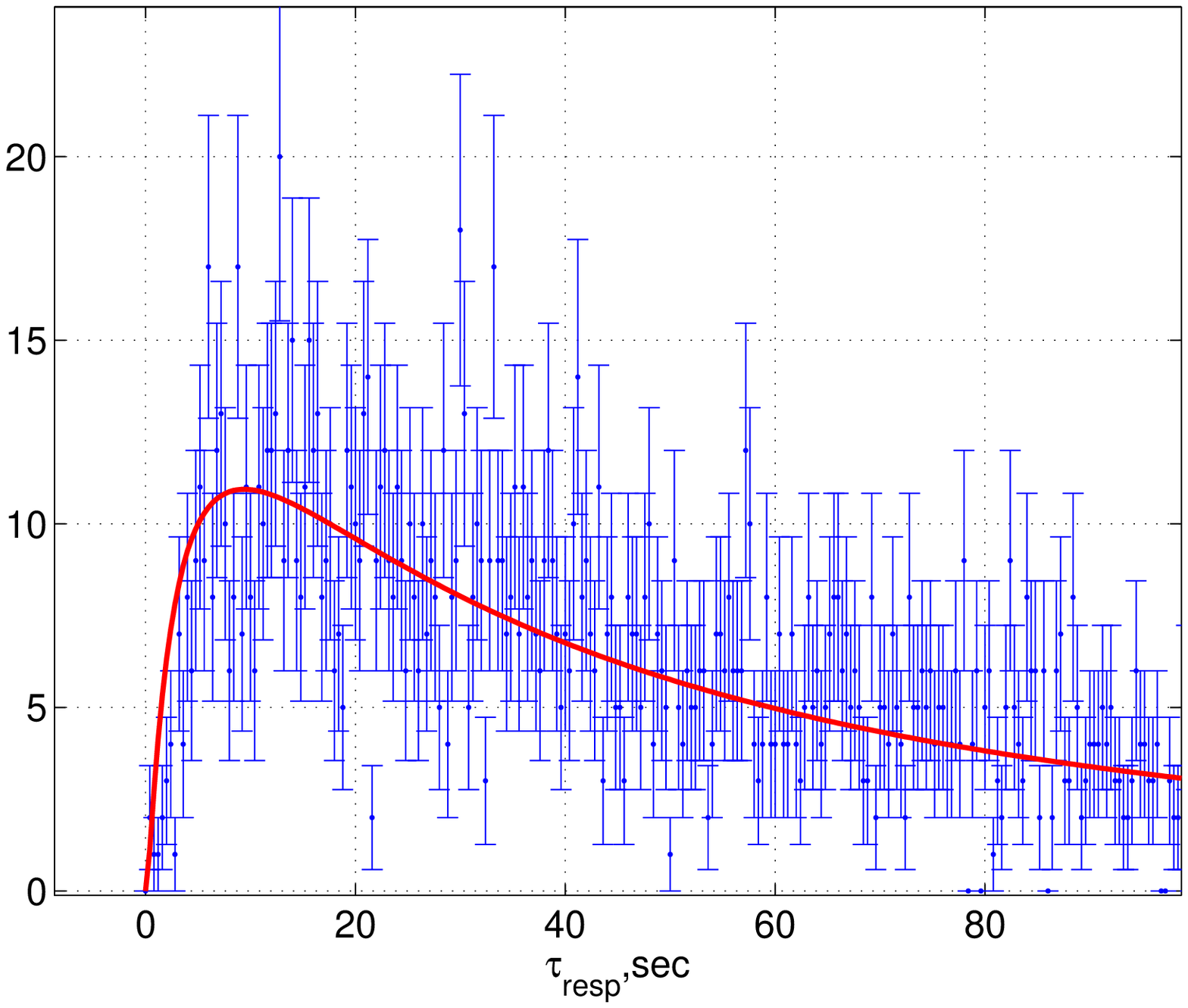} 
\end{figure}

\begin{figure}[hbp] 
\centering 
\topcaption{Quantiles of $\log{\tau_{resp}}$ of ABN-AMRO compared with quantiles of Normal distribution~(1) 
and with quantiles of $\log{\tau_{resp}}$ of AEGON/NL~(2), Daimler-Crysler/GE~(3), BBVA/SP~(4), 
Repsol YPF/SP~(5) and AXA/FR~(6).
\label{fig:quantiles}} 
\includegraphics[width=1.1\textwidth, height=0.8\textheight]{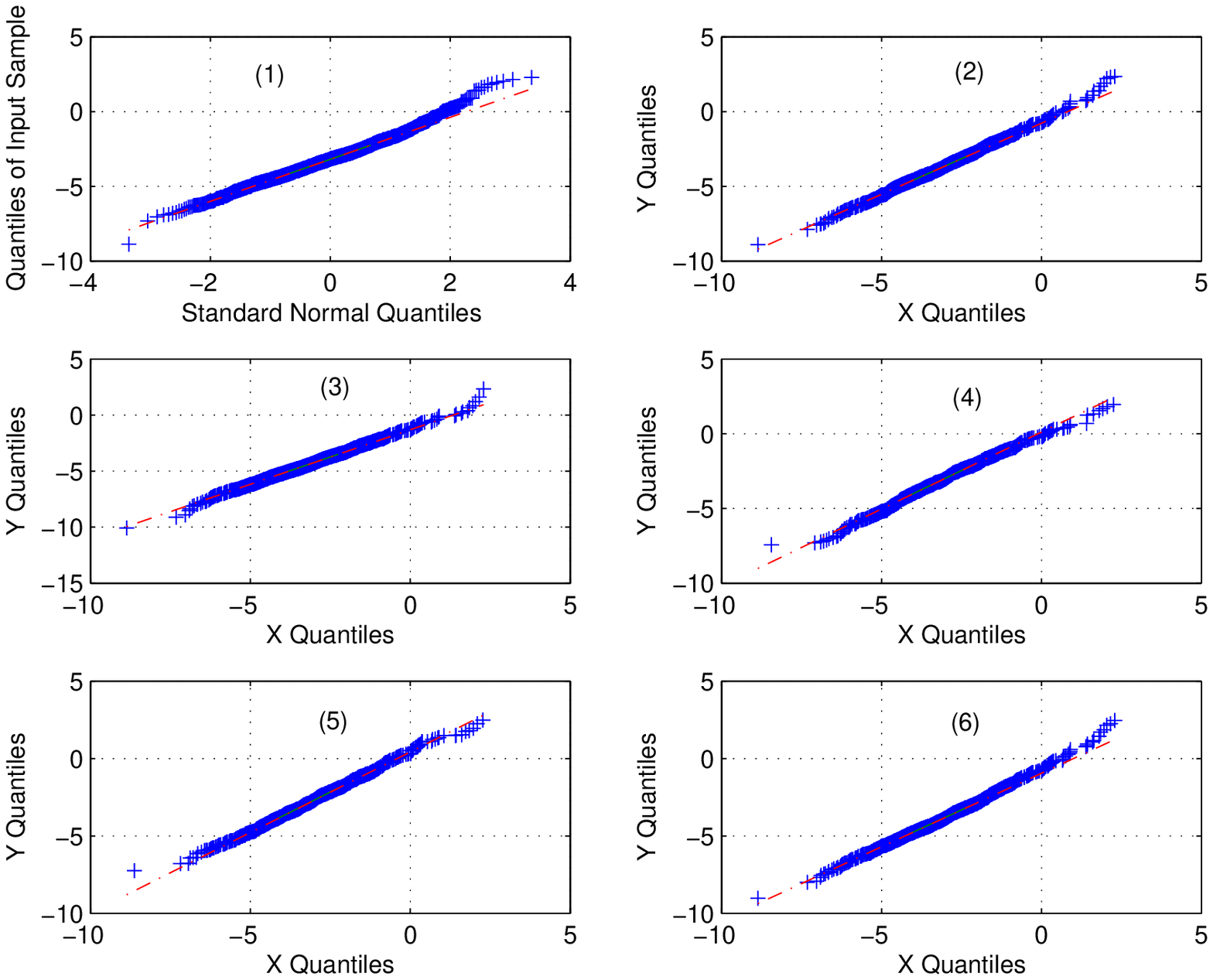} 
\end{figure}

Let us make trivial rearrangement of Equation (\ref{eq:arb3}) by moving observables 
to the left side of inequality and dividing both sides by $\mu$:
\begin{equation} \label{eq:arb4}
\frac{Spr}{|\mu|} - \tau_S \geq \tau_M + \frac{fee}{|\mu|}
\end{equation}
To test this expression we measure $Spr$ as $\tilde{Spr}$, the average update time, $\tau_S$, as $\tau_u$ 
and $\mu$ as trend slope (all are defined in previous section (sec. \ref{sec:trend_id})). 
A distribution of $\tau_{resp}=\frac{\tilde{Spr}_i}{|\mu_i|} - (\tau_u)_i$ value for all $i$-trends 
found in ABN-AMRO stock price, is plotted on Figure~\ref{fig:market_response}. 
Comparison of ABN-AMRO quantiles of $\log{\tau_{resp}}$ with quantiles of normal distribution 
and with quantiles of $\log{\tau_{resp}}$ for several other stocks indicate their similarity to 
log-normal p.d.f. and between each other, see Figure~\ref{fig:quantiles}.

\begin{figure}[hbp] 
\centering 
\topcaption{$\tau_M$ values measured per stock (numbered from 1 to 33). 
Results are sorted by exchanges where stocks are traded in bands separated 
by vertical lines. Exchanges (left to right) are Amsterdam (NLD), Frankfurt (GER), 
Madrid (SPA), Milano (ITL) and Paris (FRA). 
Horizontal lines indicate range of $\tau_M$ for each exchange. 
\label{fig:market_resolution}} 
\includegraphics[width=0.75\textwidth]{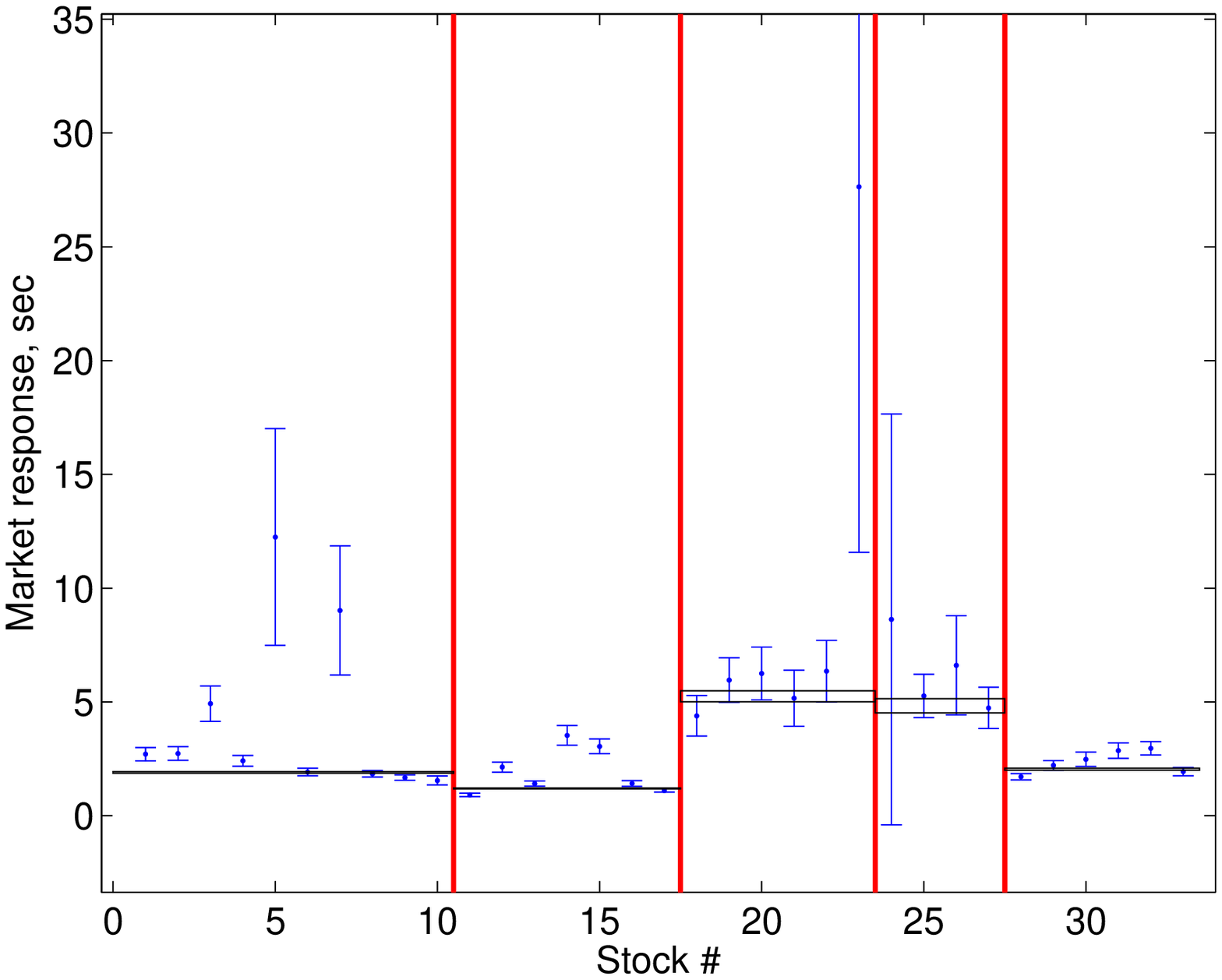} 
\end{figure}

$\tau_{resp}$-distribution starts sharply at some moment after 'zero'-time which is market reaction time, $\tau_M$ (see (\ref{eq:arb4})), 
assuming $\frac{fee}{\mu}$ equal to zero\footnote{$fee/\mu=0$ is a limit, when market-maker does not pay 
fee per transaction and $\mu$ is large enough.}. $\tau_M$ value is calibrated using $max(0, f(x))$, 
where $f(x)$ is log-normal function.
The front-end value is calculated at 1\% of fitted log-normal p.d.f.. 
$\tau_M$ values per stock are plotted in Figure \ref{fig:market_resolution}. 
Exchange averages of $\tau_M$, $\tilde{\tau_M}$, are given on the same plot and 
are quoted in Table~\ref{tab:market_resolution}. 
\begin{table}
\centering
\topcaption{Exchange averages of measured $\tau_M$. 
\label{tab:market_resolution}}
\begin{tabular}[hbp]{|c|c|}
\hline
Exchange & $\tilde{\tau_M}$ \\
\hline
\hline
Amsterdam (NLD) & 1.9 $\pm$ 0.1 \\
\hline
Frankfurt (GER) & 1.22 $\pm$ 0.04 \\
\hline
Madrid (SPA) & 5.5 $\pm$ 0.5 \\
\hline
Milano (ITA) & 5.1 $\pm$ 0.6 \\
\hline
Paris (FRA) & 2.1 $\pm$ 0.1 \\
\hline 
\end{tabular}
\end{table}

$\tau_M$ values are grouped clearly by respective exchanges, which suggest universality of 
this parameter within every particular market. In principle, the inequality~(\ref{eq:arb4}) does not force all 
Specialists to set prices close to the very limit, $\tau_M+\frac{fee}{|\mu|}$. Competition does, however.

\subsubsection{Structure of $\tau_{resp}$: $fee$ and again $\tau_M$}
\label{sec:structure}

\begin{figure}[!hbp]
\centering
\topcaption{Left plot: 2D-distribution of $\tau_{resp}$ (Y-axis) and 
$1/|\mu|$ (X-axis) for ABN-AMRO. Solid line indicates edge. 
Right plot: Edge values of $\tau_{resp}$ obtained from vertical slices of 2D-plot on the left.
\label{fig:2d_edge1}}
\includegraphics[width=\textwidth, height=0.3\textheight]{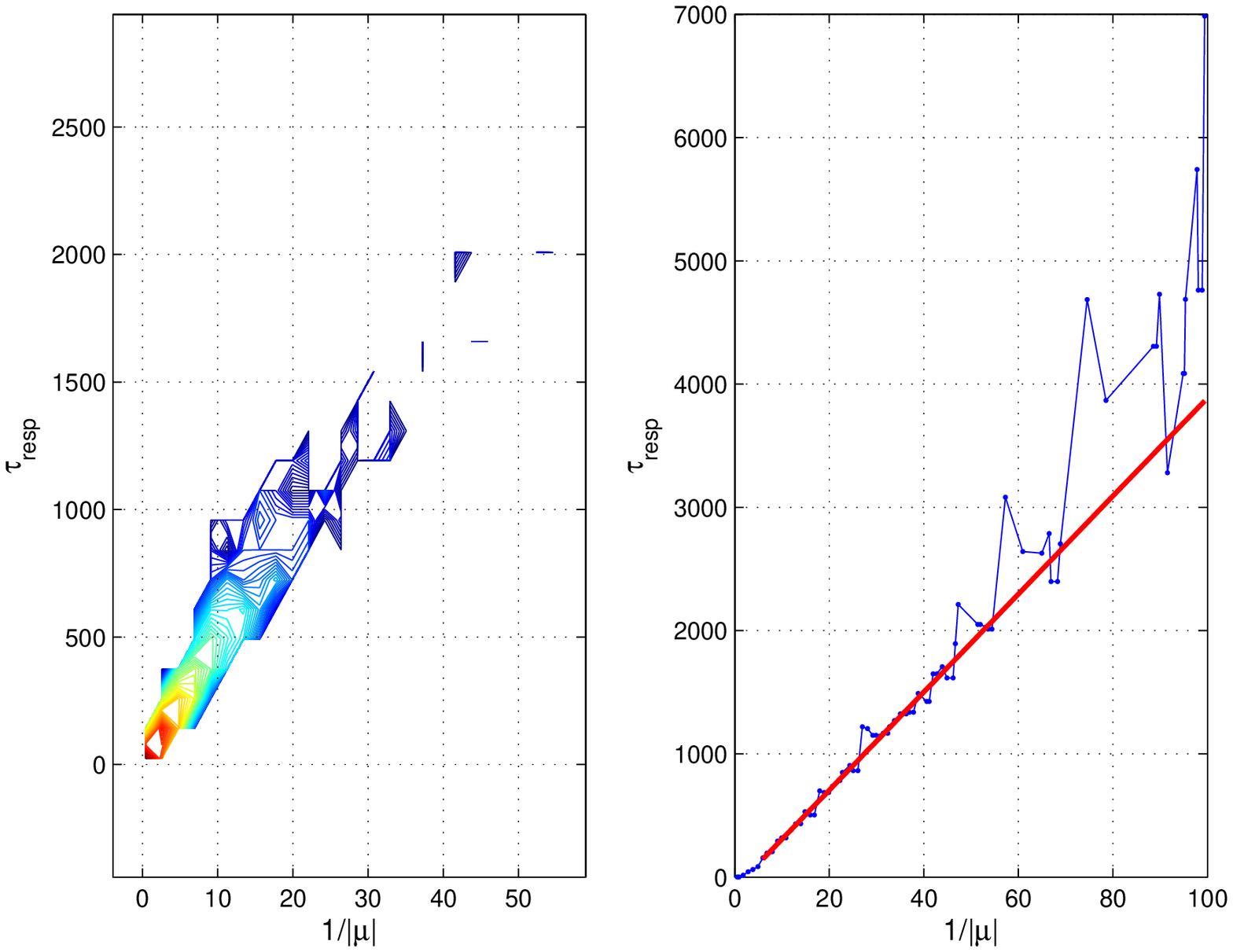}
\end{figure}

\begin{figure}[!hbp]
\centering
\topcaption{Zoom of Figure \ref{fig:2d_edge1}. Left-plot: 2D-distribution of $\tau_{resp}$ (Y-axis) and 
$1/|\mu|$ (X-axis) for ABN-AMRO. Solid line indicates edge. 
Right plot: Edge values of $\tau_{resp}$ obtained from vertical slices of 2D-plot on the left.
\label{fig:2d_edge2}}
\includegraphics[width=\textwidth, height=0.3\textheight]{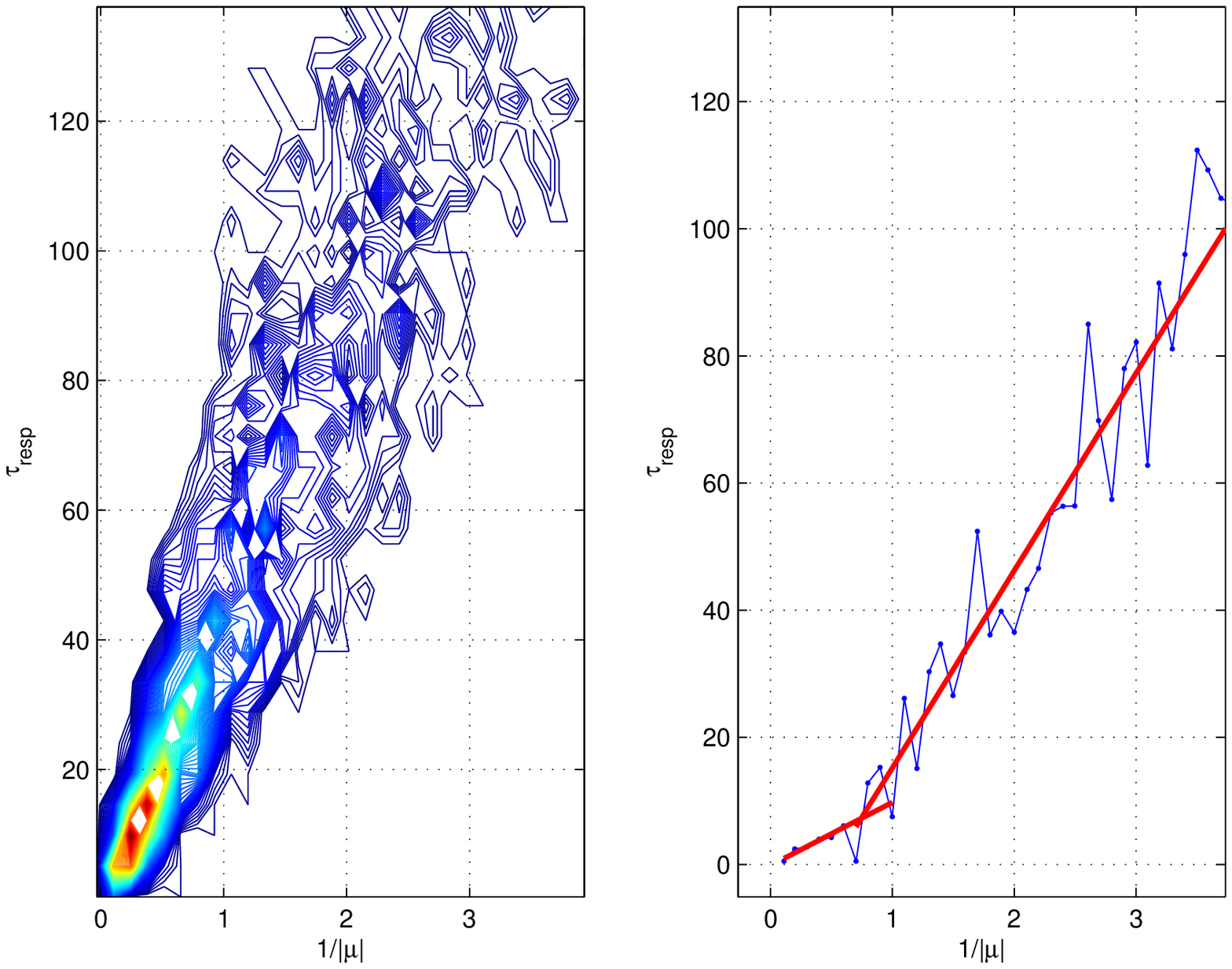}
\end{figure}

Inequality (\ref{eq:arb4}) defines linear edge $\tau_M + \frac{fee}{|\mu|}$ 
for 2D-distribution $(\tau_{resp};\frac{1}{|\mu|})$. Figure~\ref{fig:2d_edge1} supports this idea in general. 
However, zooming into the picture, see Figure~\ref{fig:2d_edge2}, suggests some structure which
probably reflects the investment horizon of Investors: fast (short/intraday horizon), medium and slow (long horizon). 

\begin{figure}[!hbp]
\centering
\topcaption{Averaged fee (X-axis) and $\tau_M$ (Y-axis). 
Three points with error bars are for (by increased $fee$ value) fast, medium and slow types of Investors. 
Averaging is done over all stocks. 
\label{fig:fee_taum}}
\includegraphics[width=0.75\textwidth]{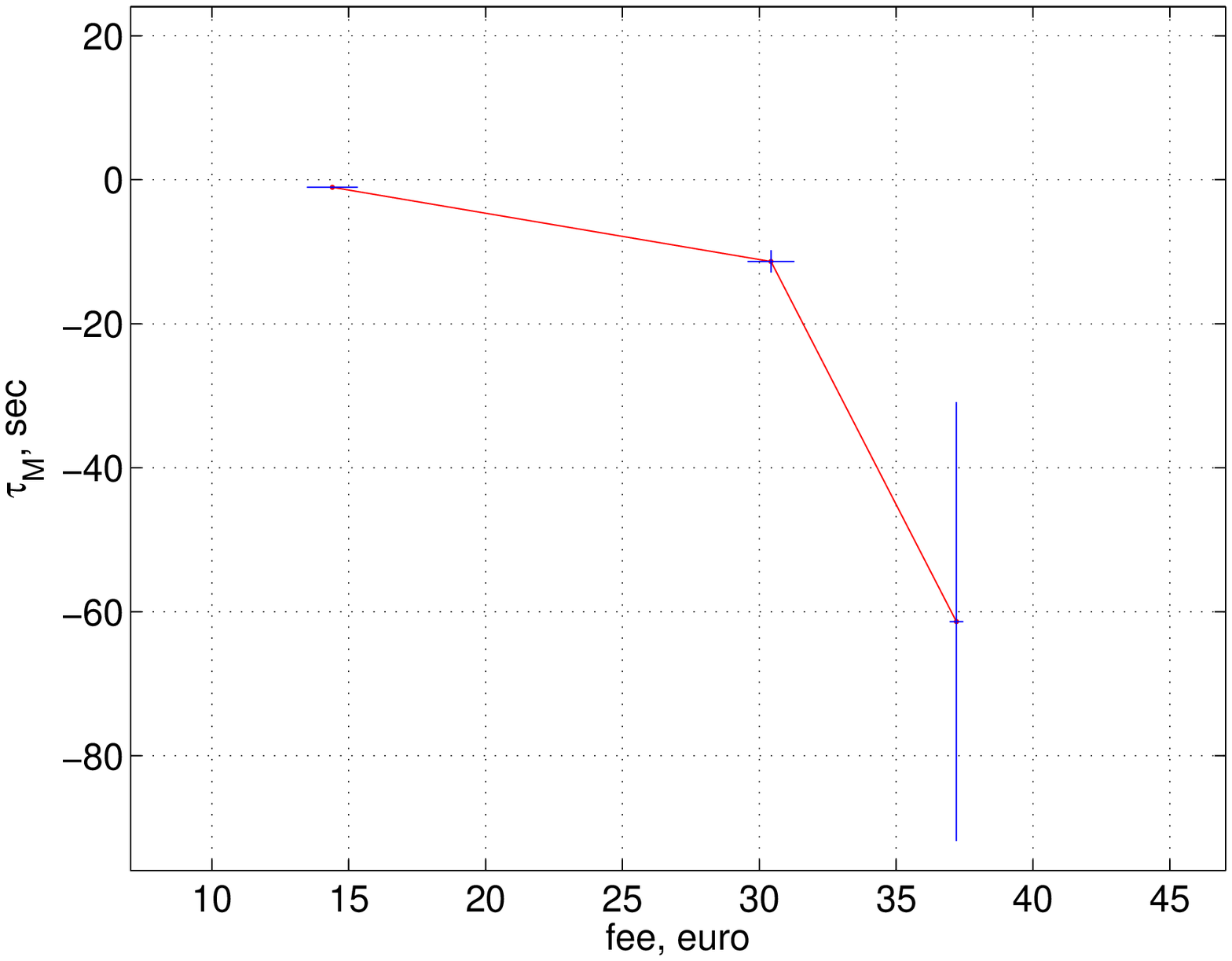}
\end{figure}

Measurement of $fee$ and $\tau_M$ parameters for all three groups for all stocks are given in the 
Table~\ref{tab:tau_m_fee}.
\begin{table}
\centering
\topcaption{$\tau_M$ and $fee$ measured from $(\tau_{resp},\frac{1}{\mu})$ 2D-distribution. 
\label{tab:tau_m_fee}}
\begin{tabular}[hbp]{|c|c|c|c|c|c|c|c|}
\hline
Market & Stock 
& \multicolumn{3}{|c|}{$fee \pm \epsilon_{fee}$}
& \multicolumn{3}{|c|}{$\tau_M \pm \epsilon_{\tau_M}$} \\
\cline{3-8}
& No. & fast & medium & slow & fast & medium & slow \\
\hline
\hline
NLD & 1 & 10 $\pm$ 2 & 30 $\pm$ 6 & 38 $\pm$ 2 & -0.1 $\pm$ 0.3 & -14 $\pm$ 11 & -62 $\pm$ 187 \\ 
 \cline{2-8}
& 2 & 14 $\pm$ 3 & 32 $\pm$ 5 & 37 $\pm$ 1 & -1.5 $\pm$ 0.4 & -17 $\pm$ 10 & -63 $\pm$ 109 \\ 
 \cline{2-8}
& 3 & 25 $\pm$ 9 & 32 $\pm$ 4 & 35 $\pm$ 1 & -2.9 $\pm$ 1.4 & -12 $\pm$ 8 & -63 $\pm$ 135 \\ 
 \cline{2-8}
& 4 & 12 $\pm$ 24 & 28 $\pm$ 3 & 38 $\pm$ 2 & -1.3 $\pm$ 2.2 & -10 $\pm$ 6 & -79 $\pm$ 278 \\ 
 \cline{2-8}
& 5 & 171 $\pm$ 48 & 190 $\pm$ 36 & 443 $\pm$ 56 & -7.3 $\pm$ 4.6 & -21 $\pm$ 63 & -1560 $\pm$ 7660 \\ 
 \cline{2-8}
& 6 & 3 $\pm$ 33 & 34 $\pm$ 4 & 38 $\pm$ 1 & -2.2 $\pm$ 3.0 & -16 $\pm$ 7 & -56 $\pm$ 84 \\ 
 \cline{2-8}
& 7 & 33 $\pm$ 15 & 88 $\pm$ 31 & 40 $\pm$ 6 & 3.8 $\pm$ 1.4 & -38 $\pm$ 53 & -36 $\pm$ 694 \\ 
 \cline{2-8}
& 8 & 12 $\pm$ 6 & 31 $\pm$ 4 & 41 $\pm$ 2 & -1.5 $\pm$ 0.6 & -14 $\pm$ 7 & -84 $\pm$ 235 \\ 
 \cline{2-8}
& 9 & 17 $\pm$ 11 & 33 $\pm$ 4 & 38 $\pm$ 1 & -4.4 $\pm$ 1.0 & -15 $\pm$ 6 & -53 $\pm$ 166 \\ 
 \cline{2-8}
& 10 & 23 $\pm$ 6 & 32 $\pm$ 5 & 52 $\pm$ 3 & -3.8 $\pm$ 0.6 & -4 $\pm$ 8 & -103 $\pm$ 408 \\ 
 \hline
 \hline
GER & 11 & 42 $\pm$ 9 & 82 $\pm$ 14 & 123 $\pm$ 16 & -3.9 $\pm$ 0.8 & -28 $\pm$ 24 & -301 $\pm$ 1993 \\ 
 \cline{2-8}
& 12 & 26 $\pm$ 7 & 37 $\pm$ 7 & 42 $\pm$ 1 & -2.6 $\pm$ 0.6 & -12 $\pm$ 12 & -51 $\pm$ 160 \\ 
 \cline{2-8}
& 13 & 20 $\pm$ 4 & 31 $\pm$ 3 & 43 $\pm$ 1 & -2.2 $\pm$ 0.3 & -6 $\pm$ 6 & -50 $\pm$ 173 \\ 
 \cline{2-8}
& 14 & 16 $\pm$ 8 & 26 $\pm$ 4 & 38 $\pm$ 1 & -1.1 $\pm$ 1.2 & -9 $\pm$ 9 & -66 $\pm$ 142 \\ 
 \cline{2-8}
& 15 & -1 $\pm$ 8 & 29 $\pm$ 4 & 38 $\pm$ 1 & 3.5 $\pm$ 0.7 & -13 $\pm$ 7 & -60 $\pm$ 125 \\ 
 \cline{2-8}
& 16 & 50 $\pm$ 17 & 75 $\pm$ 18 & 193 $\pm$ 30 & -3.9 $\pm$ 1.6 & -3 $\pm$ 32 & -546 $\pm$ 3626 \\ 
 \cline{2-8}
& 17 & 26 $\pm$ 3 & 47 $\pm$ 6 & 51 $\pm$ 3 & -2.6 $\pm$ 0.3 & -13 $\pm$ 10 & -67 $\pm$ 417 \\ 
 \hline
 \hline
SPN & 18 & -1 $\pm$ 4 & 28 $\pm$ 3 & 34 $\pm$ 1 & 3.4 $\pm$ 0.6 & -12 $\pm$ 6 & -69 $\pm$ 168 \\ 
 \cline{2-8}
& 19 & 24 $\pm$ 4 & 31 $\pm$ 7 & 37 $\pm$ 1 & -2.4 $\pm$ 0.6 & -10 $\pm$ 14 & -64 $\pm$ 182 \\ 
 \cline{2-8}
& 20 & 25 $\pm$ 13 & 27 $\pm$ 7 & 36 $\pm$ 2 & -3.3 $\pm$ 1.9 & -5 $\pm$ 13 & -53 $\pm$ 205 \\ 
 \cline{2-8}
& 21 & 19 $\pm$ 7 & 31 $\pm$ 6 & 33 $\pm$ 1 & 0.0 $\pm$ 1.7 & -15 $\pm$ 15 & -69 $\pm$ 167 \\ 
 \cline{2-8}
& 22 & 30 $\pm$ 7 & 18 $\pm$ 6 & 32 $\pm$ 1 & -4.1 $\pm$ 1.7 & -1 $\pm$ 14 & -58 $\pm$ 124 \\ 
 \cline{2-8}
& 23 & -10 $\pm$ 81 & 83 $\pm$ 33 & 43 $\pm$ 5 & 91 $\pm$ 177 & -103 $\pm$ 325 & -76 $\pm$ 582 \\ 
 \hline
 \hline
ITL & 24 & 24 $\pm$ 22 & 113 $\pm$ Inf & 36 $\pm$ 3 & 1.4 $\pm$ 69.0 & -284 $\pm$ Inf & -82 $\pm$ 388 \\ 
 \cline{2-8}
& 25 & 18 $\pm$ 7 & 25 $\pm$ 3 & 33 $\pm$ 1 & -0.2 $\pm$ 1.1 & -10 $\pm$ 6 & -56 $\pm$ 137 \\ 
 \cline{2-8}
& 26 & 29 $\pm$ 10 & 36 $\pm$ 11 & 30 $\pm$ 3 & -3.4 $\pm$ 3.1 & -20 $\pm$ 27 & -65 $\pm$ 326 \\ 
 \cline{2-8}
& 27 & 14 $\pm$ 7 & 21 $\pm$ 4 & 34 $\pm$ 1 & 0.4 $\pm$ 1.5 & -5 $\pm$ 9 & -65 $\pm$ 173 \\ 
 \hline
 \hline
FRA & 28 & 1 $\pm$ 6 & 49 $\pm$ 4 & 47 $\pm$ 3 & 0.2 $\pm$ 0.5 & -22 $\pm$ 7 & -94 $\pm$ 333 \\ 
 \cline{2-8}
& 29 & 5 $\pm$ 6 & 28 $\pm$ 14 & 39 $\pm$ 1 & 1.3 $\pm$ 0.6 & -16 $\pm$ 24 & -69 $\pm$ 123 \\ 
 \cline{2-8}
& 30 & 24 $\pm$ 11 & 41 $\pm$ 6 & 39 $\pm$ 3 & -3.0 $\pm$ 1.0 & -20 $\pm$ 10 & -34 $\pm$ 362 \\ 
 \cline{2-8}
& 31 & 16 $\pm$ 8 & 19 $\pm$ 5 & 38 $\pm$ 1 & -1.4 $\pm$ 0.8 & -3 $\pm$ 9 & -72 $\pm$ 165 \\ 
 \cline{2-8}
& 32 & 6 $\pm$ 4 & 27 $\pm$ 5 & 37 $\pm$ 1 & 1.0 $\pm$ 0.4 & -9 $\pm$ 9 & -53 $\pm$ 151 \\ 
 \cline{2-8}
& 33 & 9 $\pm$ 8 & 29 $\pm$ 4 & 40 $\pm$ 1 & -1.7 $\pm$ 0.7 & -14 $\pm$ 7 & -62 $\pm$ 147 \\ 
 \hline
 \hline
 \end{tabular}
\end{table}
Division of Investors into three groups is a reasonable approach and reflects market situation when 
the less automation, the slower their response to signals, the larger the fee (see Figure~\ref{fig:fee_taum}). 
The result also justifies our assumption that $fee=0$ when $|\mu|$ is large. $Fee$ values obtained from the fit
are too big and probably must be attributed to some other source than trading costs, however, we clearly see that
the structure of (\ref{eq:arb4}) is established correctly.

We realize that the division into groups of Investors is voluntary and is not covered 
by the model described in~(\ref{eq:arb4}). This effect should be accounted in future developments.

Concluding the test of inequality~(\ref{eq:arb5}) we see that in average market preserves the trend arbitrage 
requirements. In particular situations, however, given on page~\pageref{cond:trend_arb}, 
an arbitrage is still possible.

\subsection{Tick size and Specialists activity}

\subsubsection{Formula}

Tick size, $dp$, is another important ingredient of the trend arbitrage equation. 
Let us use similar arguments about trend arbitrage (like used in section~\ref{sec:trend_arb}, \textit{"Trend arbitrage"}).
Consider the same situation when ask prices move with up-trend (see Figure~\ref{fig:trend_arb}). 
If to assume that the price is updated every time when the expected price becomes larger 
than the current one by tick-size, the number of price updates should be:
\begin{eqnarray*} 
N_u \sim \frac{\Delta p}{dp} = \frac{|\mu|\Delta T}{dp} \textrm{ and therefore } \\
\end{eqnarray*}
\begin{equation} 
\label{eq:tick_arb}
dp \sim |\mu|\frac{\Delta T}{N_u} = |\mu|\tau_u
\end{equation}
\\where, $\Delta T$ is duration of the trend, $\Delta p$ is price change and 
$N_u$ is number of ask updates during same period and $\tau_u$ is average update time alongside the trend.
Which sign must be used in the relationship, ($=$) or strict inequality, is difficult to say because the spread 
compensates for low update frequency. Therefore we leave ($\sim$)-sign.

\subsubsection{Analysis}

\begin{figure}[!hbp]
\centering
\topcaption{Distribution of $\mu\cdot\tau_u$ for ABN-AMRO (left) with fitted Normal p.d.f.. 
Right plot is quantile plot of the same value versus normal quantile.
\label{fig:mu_tau_invariant}}
\includegraphics[width=\textwidth, height=0.3\textheight]{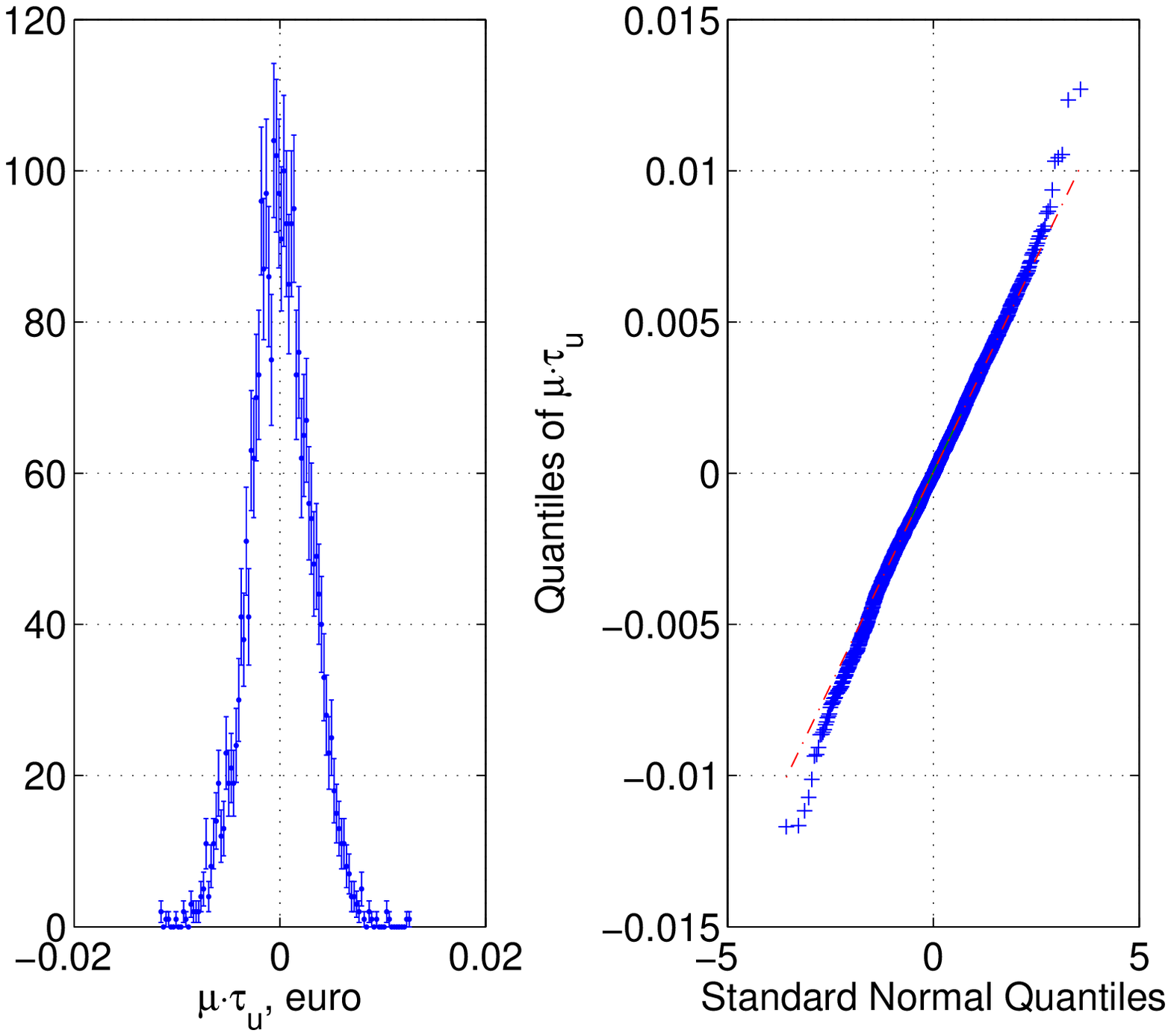}
\end{figure}

Data demonstrate, that distribution of $\mu \cdot \tau_u$ follows normal distribution with some tails, which can be 
attributed to some inefficiency in the process, see Figure~\ref{fig:mu_tau_invariant}. 
Tick size is covered within $3\sigma$ of the distribution. It means that
price update happens more often than necessary giving almost no chance for trend arbitrage. 

\begin{figure}[!hbp]
\centering
\topcaption{$\frac{3\sigma(\mu\tau)}{dp}$ of $\mu\tau$ distributions. Some anomalies of unknown origin show-up for 
two German stocks \# 11 and 16 (Allianz and SAP, respectively).
\label{fig:sigma3tick}}
\includegraphics[width=0.75\textwidth]{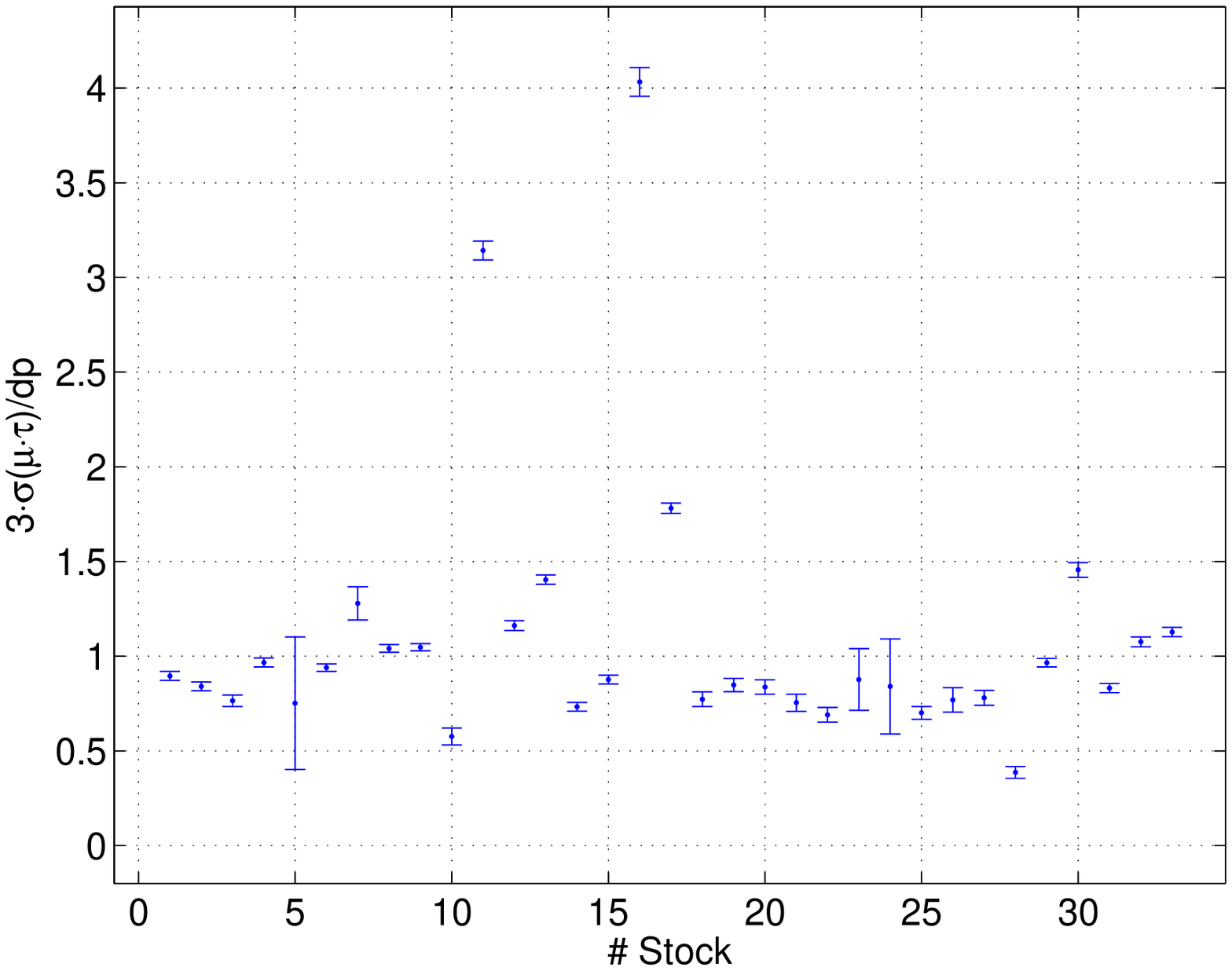}
\end{figure}

Figure~\ref{fig:sigma3tick} plots the $3\sigma(\mu\tau)/dp$ of $\mu\tau$ distributions for all stocks under 
consideration. We see that with few exceptions update frequency is such that $3\sigma$ of $\mu\tau$ is 
within tick size which confirms our proposition.

The general conclusion is that the tick size closely influences trading activity by pushing Specialists 
to update their future price view. The larger tick size, the less Specialist is motivated to do so. 
We see such pattern for some stocks of Euronext. When their price crosses 50 $euros$ 
boundary~\footnote{Euronext rules that tick size is $0.01$ $euro$ for price stock $<50$ $euro$ 
and it is $0.05$ for price stock $>50$ $euro$ and $<500$ $euro$.} the price update frequency drops accordingly.

Following trend arbitrage logic, if the tick size is small but exchange or trading systems are 
slow (e.g. Specialists trading on floor), then the $\frac{3\sigma(\mu\tau)}{dp}$ must be larger than 1. If this is
not compensated with larger spread the arbitrage exists.

\begin{figure}[hbp] 
\centering 
\topcaption{$\tau_M$ values measured per stock (numbered from 1 to 33) using~(\ref{eq:arb5}). 
Results are sorted by exchanges where stocks are traded in bands separated 
by vertical lines. Exchanges (left to right) are Amsterdam (NLD), Frankfurt (GER), 
Madrid (SPA), Milano (ITL) and Paris (FRA). 
Horizontal lines indicate range of $\tau_M$ for each exchange. 
\label{fig:market_resolution2}} 
\includegraphics[width=0.75\textwidth]{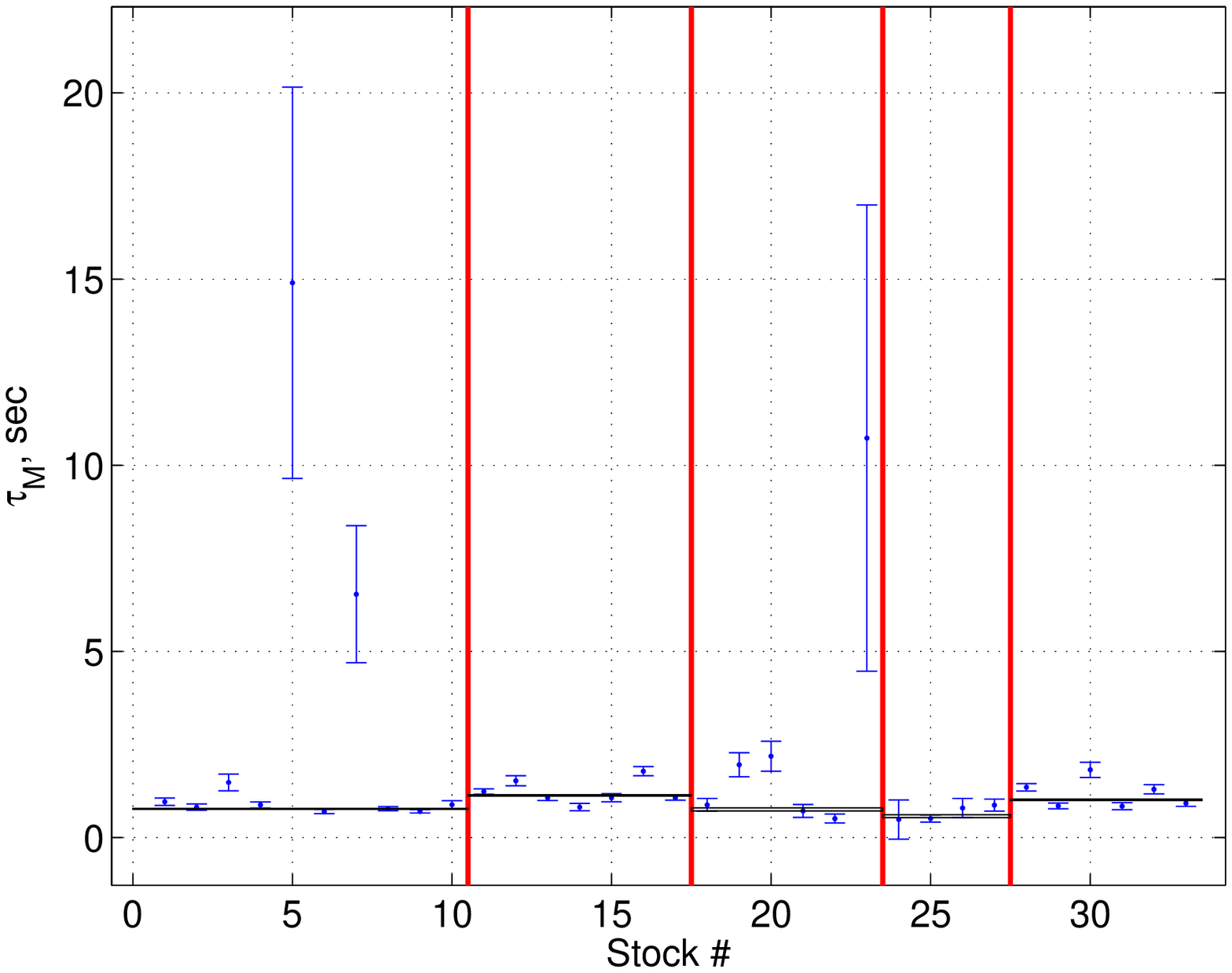} 
\end{figure}

\begin{figure}[!hbp]
\centering
\topcaption{Averaged fee (X-axis) and $\tau_M$ (Y-axis). 
Three points with error bars are for (by increased $fee$ value) slow, medium and fast types of Investors. 
Averaging is done over all stocks.
\label{fig:fee_taum2}}
\includegraphics[width=0.75\textwidth]{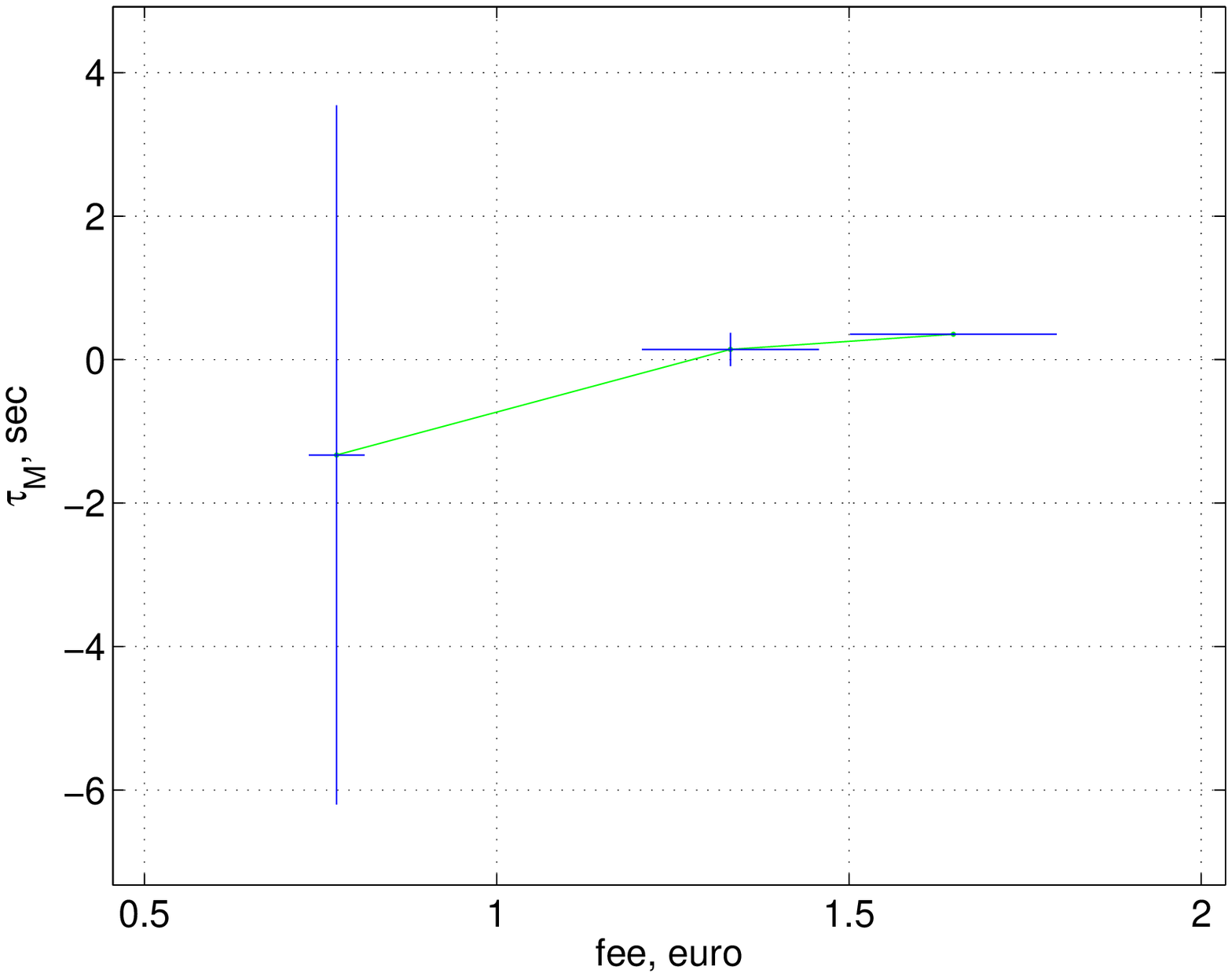}
\end{figure}

By substituting (\ref{eq:tick_arb}) into (\ref{eq:arb4}) we obtain the following:
\begin{equation} \label{eq:arb5}
\frac{Spr-dp}{|\mu|} \geq \tau_M + \frac{fee}{|\mu|}
\end{equation}
Performing the same measurement of $\tau_M$ as in section \ref{sec:market_reaction} we find that the distinction between 
exchanges is almost removed and is now ranges between 0.6 and 1.2 second (see Figure~\ref{fig:market_resolution2} 
and Table~\ref{tab:market_resolution2}. Taking $dp=0.01$ $eur$ (and $0.05$ $eur$ for some stocks of Euronext) and fitting edge 
of ($\frac{Spr-dp}{|\mu|}$,$\frac{1}{|\mu|})$-distribution as it was done before 
in section~\ref{sec:structure} reveals (see Figure~\ref{fig:fee_taum2}) that tick size, $dp$, absorbs both
$\tau_S$ and (probably) $\tau_M$, therefore, reducing (\ref{eq:arb5}) to the following:
\begin{equation} \label{eq:arb6}
Spr-dp \geq \tau_M\cdot\mu + fee
\end{equation}

\begin{table}
\centering
\topcaption{Exchange averages of measured $\tau_M$.  Eq~(\ref{eq:arb5}) is used.
\label{tab:market_resolution2}}
\begin{tabular}[hbp]{|c|c|}
\hline
Exchange & $\tilde{\tau_M}$ \\
\hline
\hline
Amsterdam (NLD) & 0.79 $\pm$ 0.03 \\
\hline
Frankfurt (GER) & 1.15 $\pm$ 0.03 \\
\hline
Madrid (SPA) & 0.8 $\pm$ 0.1 \\
\hline
Milano (ITA) & 0.6 $\pm$ 0.1 \\
\hline
Paris (FRA) & 1.04 $\pm$ 0.04 \\
\hline 
\end{tabular}
\end{table}

We may conclude from this result, that (\ref{eq:tick_arb}) absorbs all stock specific and (probably) exchange-wide 
latencies information. In fact, all results described in this article should be confirmed from actual hardware response, 
i.e. direct measurement of all latencies within exchange and traders computers/software.

\subsubsection{Simulation: Trends in Random walk model and return-trend duality}

A simplified model of price evolution has been implemented to understand empirical results. 
The price increment of this model follows $dS = \sigma dW$.
Trend search algorithm (given in section \ref{sec:trend_id}) is imposed to generated price time series.
Trends found with that procedure are stored in the same fashion as we did for real data.

Two important points have to be mentioned:
\begin{itemize}
\item The model assumes generation of prices at equal time intervals, i.e. trading activity is not simulated and
\item Spread is assumed to be constant and is not dependent on the trend;
\end{itemize}

In particular, the model reproduces Normal distribution of $|\mu|\tau_u$ for all tracks. This leads to an idea that 
the description of price evolution with some random walk model is equivalent to description of price via trends, 
which also can be stochastic~\footnote{Such model to be built.}.

To be more specific let us consider some stochastic model describing price dynamics, like Black-Scholes model, 
$BS(\mu,\sigma;\epsilon)$ with growth rate, $\mu$, volatility, $\sigma$, and $\epsilon$, some value to cut the trend, 
like $n$ in (\ref{eq:trigger}), 
as parameters. Then, there is a homeomorphic transformation in parametric space into another model 
describing same price dynamics in terms of trends, $T(\mu, \Delta T; q)$, with trend slope, $\mu$, 
trend duration, $\Delta T$, and quality, $q$, as parameters. Quality, $q$, characterizes amount of fluctuations 
around trend (spread dynamics). 

\subsubsection{Beta}

\begin{figure}[!hbp]
\centering
\topcaption{Dependence of $\tau_u$ on $\frac{\tilde{Spr}}{\mu}$ for ABN-AMRO. 
Solid line indicates $x^{0.625}$ function.
\label{fig:strong_rel_2d}}
\includegraphics[width=0.75\textwidth]{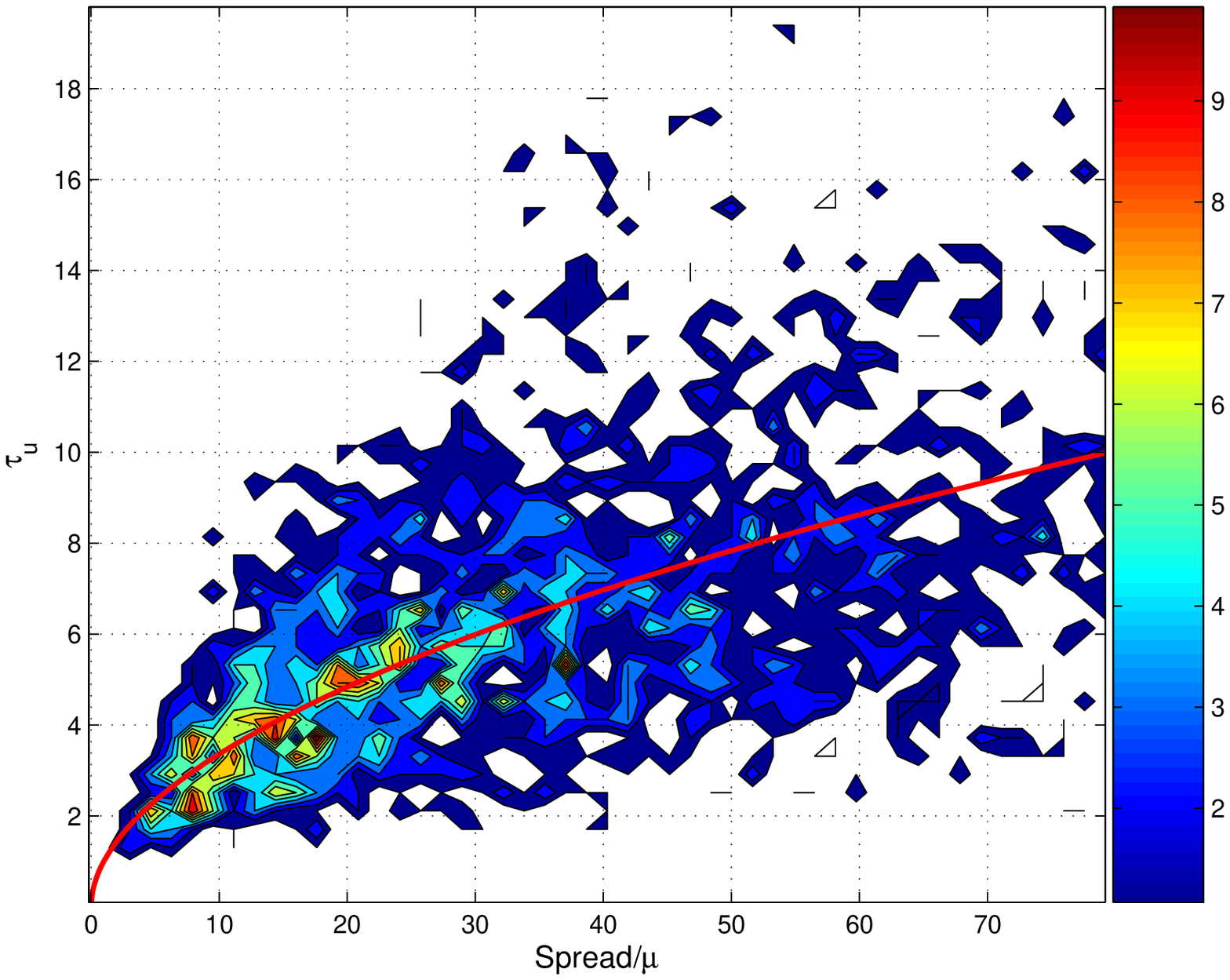}
\end{figure}

\begin{figure}[!hbp]
\centering
\topcaption{$\beta$ for all stocks. 
\label{fig:beta}}
\includegraphics[width=0.75\textwidth]{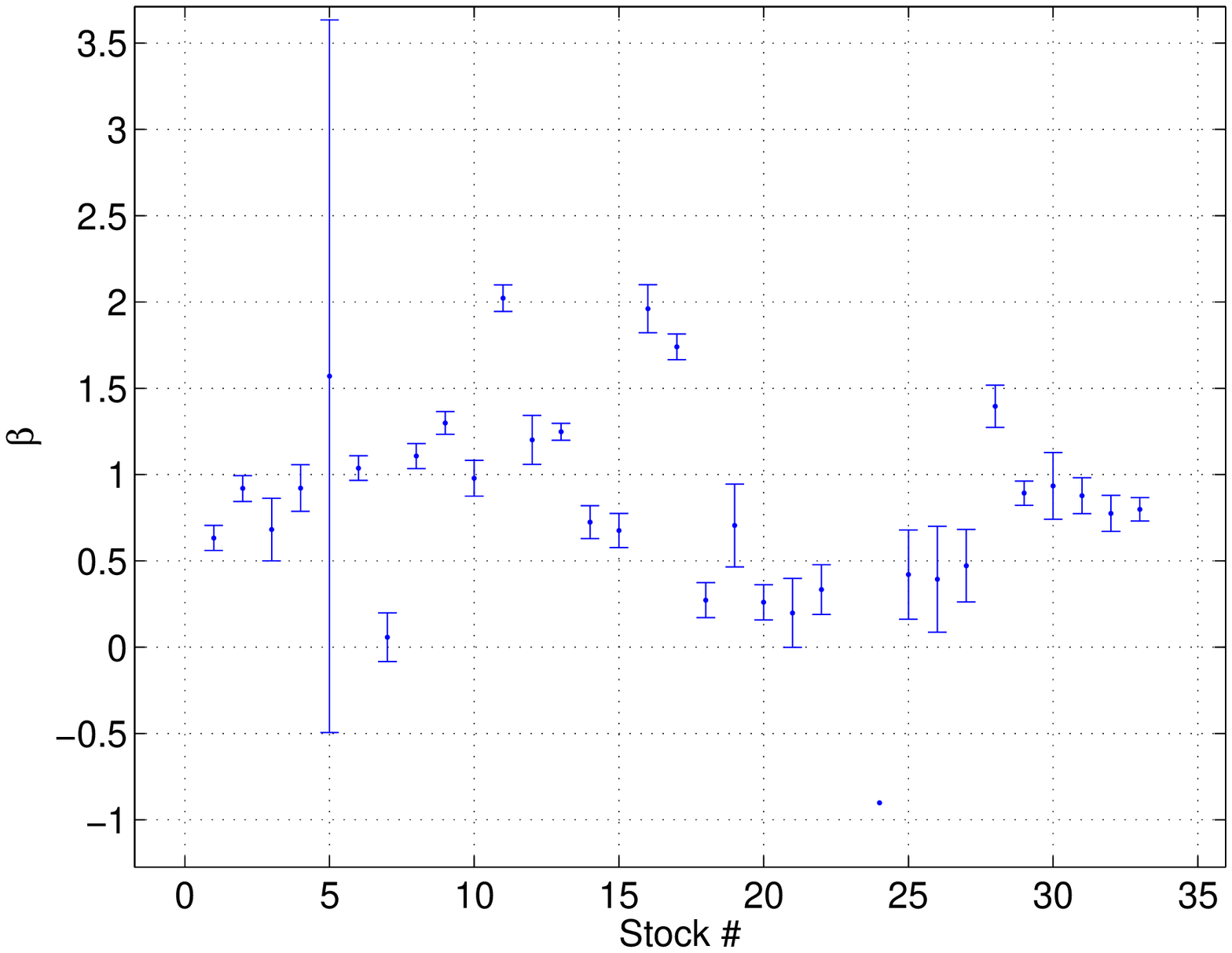}
\end{figure}

Looking closer to relationship between $\frac{\tilde{Spr}}{\mu}$ and $\tau_u = \frac{\Delta T}{N_u}$, 
Figure~\ref{fig:strong_rel_2d}, 
we can find strong relationship:
\begin{equation} \label{eq:strong_eq}
\frac{\tilde{Spr}}{\mu} = \tau_u^{1+\beta} + \eta, 
\end{equation}
\\where $\beta$ is stock dependent and ranges between 0.3 and 1.1, see Figure~\ref{fig:beta}, and $\eta$ is 
random Wiener process scaled roughly as $\tau$.
\\Indeed, the distribution of difference $\frac{\tilde{Spr}}{\mu} - \tau_u^{1+\beta}$ 
has a peak around zero and fits to Normal p.d.f. where part of distribution is cut due to trend arbitrage, 
which plays role of natural constrain for price generaion process (\ref{eq:strong_eq}). 
$\beta$ is found through two-step procedure:
\begin{itemize}
\item slices in $\frac{Spr}{|\mu|}$ are fit to Normal p.d.f., $N(m_{\tau},s)$;
\item $m$ are fit to $\frac{Spr}{|\mu|}=m_{\tau}^{1+\beta}$
\end{itemize}

It is difficult to interpret meaning of this relationship, because dimensionalities of 
l.h.s and r.h.s. of equation (\ref{eq:strong_eq}), [$sec$] and [$sec^{\beta}$], are different. 
There is some reference model though \cite{SABR}, which suggests fractionality of return process:
\begin{equation}
dF = aF^\beta dW_1; \textrm{ } da = \nu a dW_2; \textrm{ } dW_1 dW_2 = \rho dt
\end{equation}

The main difference, however, with the mentioned result is that SABR-model describes process of last prices, while 
equation (\ref{eq:strong_eq}) describes bid-ask update. The relationship between bid-ask update process and trade process 
is subject of the next article. 

\section{Conclusions and prospects}

The presented analysis has demonstrated importance of analysis of short-term trends as 
it gives better insight into micro-dynamics of the market. 

Using simple mechanistic arguments the trend arbitrage inequality is developed. 
Empirical tests prove the inequality to hold.
The inequality sets limit on bid-ask spread which is determined by the latencies of exchange and trading systems
and by some costs. These latencies can also be identified directly by measuring delay between sending time 
of the order from traders computer and appearance of this order in the exchange book.

Using same arguments, the tick size is related to frequency of price update by Specialist.
Analysis of trend inequality with data allowed us to measure latencies of exchange 

Flat distribution of probability of $\chi^2_t$ of trend fit demonstrates 
that the spread bears instant information about price uncertainty 
in the process of price measurement. Indeed, from the point of view of Investor the next \textit{last 
price} will be within the current bid-ask spread. 

Structure of trend arbitrage is in general captured correctly. 
In particular, it allows to see division of market participants by their speed. 

Further analysis suggests some strong relationship between spread, trend and time of trades update. 
Although dimensionality problem of the equation remains, the theoretical explanation of this phenomenon 
is still waiting to appear.

The final conclusion is that the price evolution can be equally described through random walk or trends.
However, market participants induce such spread dynamics as the trend exists and they limit their actions
to conform trend arbitrage requirement.

In the future research, we plan to investigate relationship between \textit{price update} and $trading$ 
processes using the same type of analysis of trend associated information. Some model development 
will take place as well.

\end{document}